\documentclass[12pt]{iopart}

\expandafter\let\csname equation*\endcsname\relax
\expandafter\let\csname endequation*\endcsname\relax

\usepackage{amssymb,amsmath,amsbsy,bm,amscd}

\usepackage{graphicx,color}
\DeclareGraphicsExtensions{.png,.pdf}

\usepackage[singlelinecheck=false,justification=justified]{caption}

\usepackage{url}
\usepackage[breaklinks=true,colorlinks=true,linkcolor=blue,urlcolor=blue,citecolor=blue]{hyperref}

\newcommand{\ket}[1]{\left| {#1} \right\rangle}
\newcommand{\bra}[1]{\left\langle{#1} \right|}

\begin{document}
\title{Cutoff phenomenon and entropic uncertainty for random quantum circuits}
\author{Sangchul Oh, Sabre Kais$^{\dag}$}
\address{Department of Chemistry, Department of Physics and Astronomy, and Purdue Quantum 
              Science and Engineering Institute, Purdue University, West Lafayette, IN, USA}
\ead{$^{\dag}$\;Corresponding author:\;kais@purdue.edu}
\vspace{10pt}
\begin{indented}
\item[]\today
\end{indented}

\begin{abstract}
How fast a state of a system converges to a stationary state is one of the fundamental questions in science. 
Some Markov chains and random walks on finite groups are known to exhibit the non-asymptotic convergence 
to a stationary distribution, called the cutoff phenomenon. Here, we examine how quickly a random quantum 
circuit could transform a quantum state to a Haar-measure random quantum state. We find that random quantum 
states, as stationary states of random walks on a unitary group, are invariant under the quantum Fourier 
transform. Thus the entropic uncertainty of random quantum states has balanced Shannon entropies for the 
computational bases and the quantum Fourier transform bases. By calculating the Shannon entropy for random 
quantum states and the Wasserstein distances for the eigenvalues of random quantum circuits, we show that 
the cutoff phenomenon occurs for the random quantum circuit. It is also demonstrated that the Dyson-Brownian 
motion for the eigenvalues of a random unitary matrix as a continuous random walk exhibits the cutoff 
phenomenon. The results here imply that random quantum states could be generated with shallow random circuits.
\end{abstract}

\vspace{2pc}
\noindent{
\submitto{{Electronic Structures: Focus Issue on ``Quantum Chemistry in the Ages ..."}}
 
%

\section{Introduction}
How many shuffles are enough to ensure that a deck of 52 cards is mixed randomly? The answer is 
$\tfrac{3}{2}\log_2 52$, approximately 8.55, for the riffle shuffling case, as shown by Aldous and 
Diaconis~\cite{Aldous_Diaconis1986} and by Bayer and Diaconis~\cite{Bayer_Diaconis1992}. Fewer than this 
number are not enough to mix the deck of cards and more do not significantly improve the randomness. This 
non-asymptotic convergence to a steady or equilibrium state is called the cutoff phenomenon~\cite{Diaconis1996} 
and has been discovered in various fields. Some finite Markov chains exhibit the cutoff phenomenon. These 
include the Ehrenfest urn model as a simplified diffusion model~\cite{Ehrenfest1907}, 
random walks on a hypercube~\cite{Diaconis1990}, 
some Metropolis algorithms~\cite{Diaconis1996}, Glauber dynamics of Ising models~\cite{Lubetzky2013}, and 
certain quantum Markov chains~\cite{Kastoryano2012}.

The card shuffling problem is considered a random walk on the symmetric group $S_{52}$. The cutoff phenomenon 
occurs also for random walks on a compact Lie group. Rosenthal~\cite{Rosenthal1994} considered that random 
walk on the special orthogonal group ${\rm SO}(N)$ given by repeated rotations by a fixed angle through 
randomly chosen planes. Rosenthal showed that the random walk on ${\rm SO}(N)$, after 
$1/(2(1-\cos\theta)) N\log N + cN$ rotations with a fixed angle $\theta$ and a constant $c$, converges rapidly 
to the Haar measure in total variation distance. Porod~\cite{Porod1996a,Porod1996b} showed that random walks 
on ${\rm O}(N)$, ${\rm U}(N)$, and ${\rm Sp}(N)$ given by random reflections converge to Haar measure in 
the total variation distances and the cutoff phenomena occurs at $\frac{1}{2} N \log N$ steps.

Random circuit sampling is a task to sample bit-strings from a probability distribution defined by random 
quantum circuits, and considered a good candidate to demonstrate quantum advantage with noisy-intermediate 
scale quantum computers. In 2019, it was implemented on the Sycamore processor with 53 qubits~\cite{Arute2019} 
and recently on the Zuchongzhi processor with 56 and 60 qubits~\cite{Wu2021,Zhu2022_60qubit}. In both quantum 
computations, the random circuit was implemented by applying repeatedly, up to 20 and 24 cycles, 
randomly-chosen single qubit gates and the two-qubit gate acting on the nearest neighbor qubits. The number 
of cycles plays the same role as the number of times a deck of cards is shuffled or the number of random 
rotations on Lie groups. So it is natural to ask the same questions which have been answered for 
the card shuffling problem. How many depths of random quantum circuits, i.e., the number of cycles, are 
needed to obtain a Haar-measure random unitary operator or to transform an initial quantum state into a pure 
random quantum state? Does the cutoff phenomenon occur? What is going on in a system after a steady state 
has been reached? Is there any tool to measure the closeness to a steady state in addition 
to the total variance distance? In this paper, we present partial answers to these questions
using the random circuit implemented on the Sycamore processor and a time-dependent random Hamiltonian 
model. The former is considered a discrete random walk on a unitary group and the latter a continuous random 
walk called the Dyson-Brownian motion for eigenvalues of a unitary operator. A pure random quantum state at 
the steady state will be analyzed with the Shannon entropy.  Instead of the total variation distance 
$||\mu_k -\mu_{\rm Haar}||_{\rm TV}$ between the probability distribution $\mu_k$ at the $k$-th time step and 
the Haar measure distribution $\mu_{\rm Haar}$, we employ the Wasserstein distance for the distribution of 
eigenvalues of a unitary operator.

The paper is organized as follows. In Sec.~\ref{Sec2}, we will discuss the cutoff phenomenon for random 
quantum circuits by calculating the Shannon entropy and the Wasserstein distance. We show that the Shannon 
entropy of a random quantum state is invariant under the quantum Fourier transform. We discuss the entropic 
uncertainty relation of random quantum states for the computational bases and the quantum Fourier transform 
bases. In Sec.~\ref{Sec3}, we investigate the cutoff phenomenon of the Dyson-Brownian motion generated by
a time-dependent random Hamiltonian. Finally, in Sec.~\ref{Sec_summary}, we will summarize the result and 
present the discussion.

\section{Cutoff Phenomenon for Random Quantum Circuits \label{Sec2}}

Let us begin with the introduction to the random quantum sampling implemented on the Sycamore quantum 
processor and on the Zuchongzhi quantum processor~\cite{Arute2019,Wu2021}. The task is to sample 
bit-strings $x = a_1a_2\cdots a_{n} \in \{0,1\}^n$ from the probability $p(x) = |\bra{x}U\ket{0}^{\otimes n}|^2$ 
given by a random quantum circuit $U$ acting on $n$ qubits where $\ket{0}^{\otimes n} = \ket{0_1\cdots0_{n}}$ 
is the initial state and $\ket{x} \equiv \ket{a_1a_2\cdots a_{n}}$ is a computational basis. The random 
quantum circuit $U$ implemented on both the Sycamore and Zuchongzhi processors is given by repeatedly 
applying random unitary operators $U_k$ and finally the single-qubit gates $S$ before the measurement 
\begin{equation}
U = SU_m \cdots U_{2} U_1 \,,
\label{Eq1}
\end{equation}
where each random quantum circuit $U_k$ is composed of single-qubit gates $S$ chosen randomly from 
the set $\{\sqrt{X},\sqrt{Y},\sqrt{W}\}$ on all qubits and two-qubit gates on the pair of qubits 
selected in the sequence of the coupler activation patterns of a 2-dimensional array of qubits. 
Millions of bit-strings were sampled from the Sycamore and Zuchongzhi processors. The distributions of 
bit strings obtained from these noisy quantum processors are deviated from the ideal distribution. 
Oh and Kais~\cite{Oh2022_JPCL,Oh2022_PRA,Oh2023_PRA}} investigated this deviation using the random matrix
theory and the Wasserstein distances.

Eq.~(\ref{Eq1}) is considered random walks or random rotations on a unitary group. If the number of cycles 
$k$ is large, $U^{(k)} \equiv U_k\cdots U_1$ would approach a random unitary operator $U_{\rm Haar}$ sampled 
from the Haar probability measure on a unitary group ${\rm U}(2^n)$. Typically, the convergence 
to a stationary state is measured by the total variation distance $||v^{*(k)} - \mu_{\rm Haar}||_{\rm TV}$ where 
$v^{*(k)}$ is the distribution of the random walk at step $k$ and $\mu_{\rm Haar}$ is the Haar 
measure~\cite{Rosenthal1994,Porod1996a,Porod1996b}. For the Sycamore processor, the sub-linear convergence, 
the depth proportional to $\sqrt{n}$, was claimed by calculating the average entropy of random quantum 
states~\cite{Boixo2018}. Emerson {\it et al.}~\cite{Emerson2003} studied a pseudo-random circuit given by 
repeated applications of $n$ single-qubit random gates sampled from the Haar measure on ${\rm U}(2)$ and 
simultaneous two-qubit interactions. Moreover, Emerson {\it et al.} employed the measure of entanglement for 
a multipartite system as the measure of the convergence. It was shown that this random quantum circuit 
converges to the Haar measure if the circuit depth $m$ is larger than $m_c\equiv {\cal O} (n^3 N^2)$ with 
$N=2^n$. This is larger than the cutoff step $k_c \equiv \frac{1}{2}N\ln N = {\cal O}(nN)$ for random 
rotations or random reflections on Lie groups shown by Rosenthal~\cite{Rosenthal1994} and 
Porod~\cite{Porod1996a,Porod1996b}.

Different measures have been used to quantify the convergence to the Haar measure distribution, and
give rise to the different cutoff steps. Here, we employ the Shannon entropy for quantum states and the 
Wasserstein distance between the eigenvalue distribution of a random unitary operator sampled from the Haar measure
and those of random quantum circuits. Random unitary matrices drawn from the Haar measure on a unitary group
are called the circular unitary ensemble. The properties of random quantum states and the eigenvalue 
distribution of the circular unitary ensemble are well known from the random matrix theory~\cite{Haake2010}.

Let us first investigate how close a quantum state at the $k$-th step, 
$\ket{\psi^{(k)}} = R^{(k)}\ket{0} = U_k\cdots U_1\ket{0}$, is to a random quantum state, a stationary
state of random walks on a unitary group. An immediate question is what a pure random quantum state is
and how to generate it. A random pure state can be generated in several ways, that is, there are several 
ways of drawing a random unitary operator from the Haar measure~\cite{Mezzadri2007,Meckes2019,Ozols2009}.
A random unitary operator could be sampled from the Haar measure through the Euler angle method or 
the QR decomposition of a complex Gaussian random matrix. Basically, a random quantum state 
$\ket{\psi} = \sum_{i=0}^{N-1} a_i\ket{i}$ can be viewed as a random vector on the $(2N-1)$ sphere 
and expansion coefficients $a_i = x_i + iy_i$ are drawn from the normal distributions, i.e., 
$x_i, y_i \sim {\cal N}(0,1)$. The distribution $P=\{p_1,p_2,\dots, p_{N-1}\}$ of probabilities 
$p_i = |a_i|^2$ of a random quantum state obeys the $\chi^2$ distribution with 2 degree of 
freedom~\cite{Haake2010} 
\begin{equation}
{\rm Pr}(p) = (N-1)(1-p)^{N-2} \,.
\label{chi2}
\end{equation}

The $\chi^2$ distribution for $P$ of $p_i$ of random quantum states makes it possible to calculate 
the average Shannon entropy. The Shannon entropy for the probability distribution $P=\{p_1,p_2,\dots,p_{N-1}\}$
is 
\begin{align}
H(P) = H(\ket{\psi}) = -\sum_{i=0}^{N-1} p_i \ln p_i\,,
\label{Shannon}
\end{align}
where $\sum_i p_i = 1$ and $p_i=|a_i|^2$. The Shannon entropy $H(\ket{\psi})$ for a quantum state measures 
the amount of 
uncertainty or the concentration of $P=(p_0,\dots,p_{N-1})$. The average Shannon entropy over random quantum 
states can be calculated with Eqs.~{(\ref{chi2})} and~{(\ref{Shannon})} and is given by
\begin{equation}
\langle H(\ket{\psi})\rangle = \ln N - 1 +\gamma\,,
\end{equation} 
where $\gamma\approx 0.5772$ is the Euler constant and $\langle \cdots\rangle$ is the average over random quantum
states. So the Shannon entropy of a quantum state could be used as a measure of convergence of random walks
on a unitary group.

\begin{figure}[t]
\centering{
\includegraphics[width=0.48\textwidth]{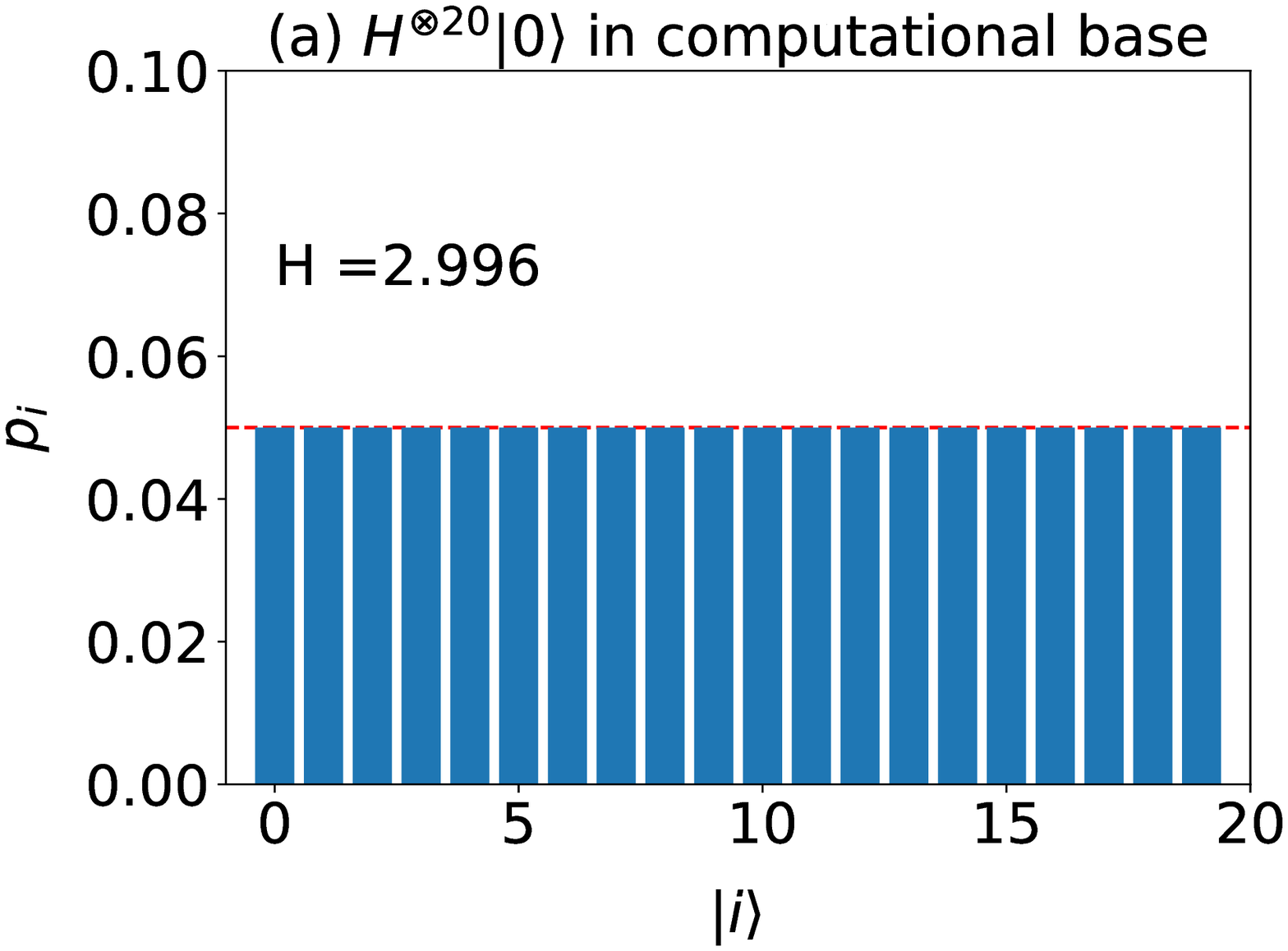}\quad
\includegraphics[width=0.48\textwidth]{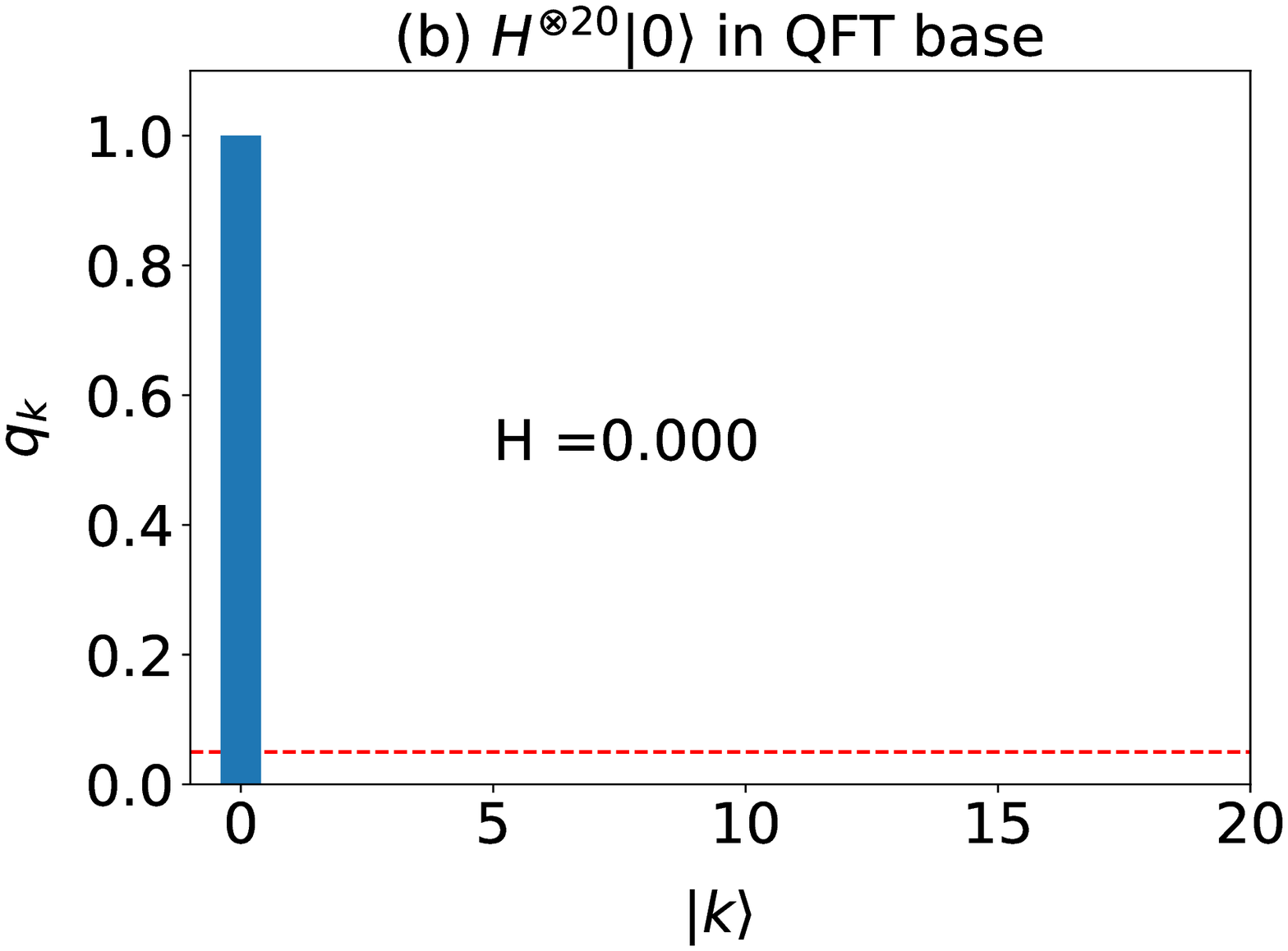}\\[10pt]
\includegraphics[width=0.48\textwidth]{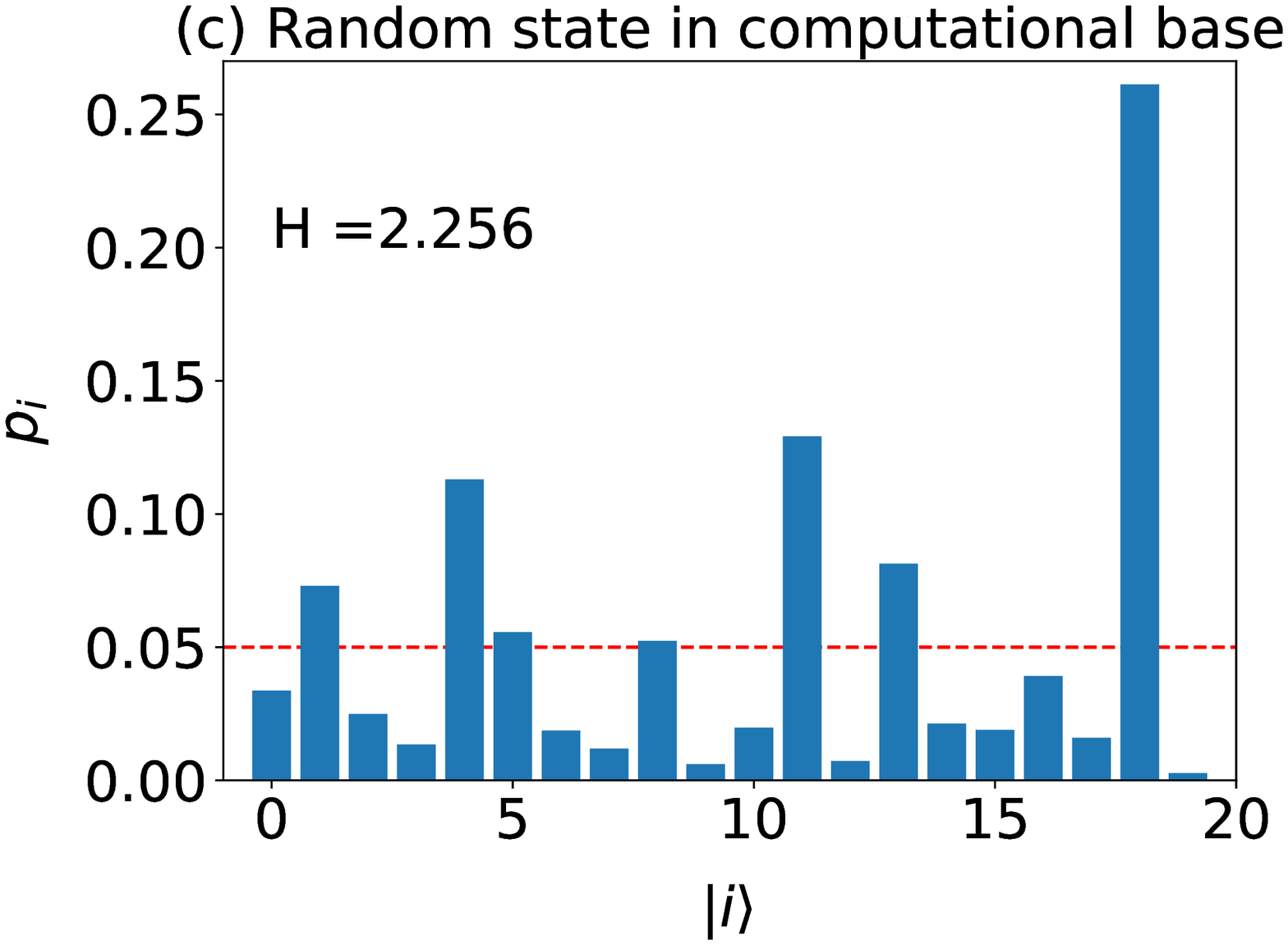}\quad
\includegraphics[width=0.48\textwidth]{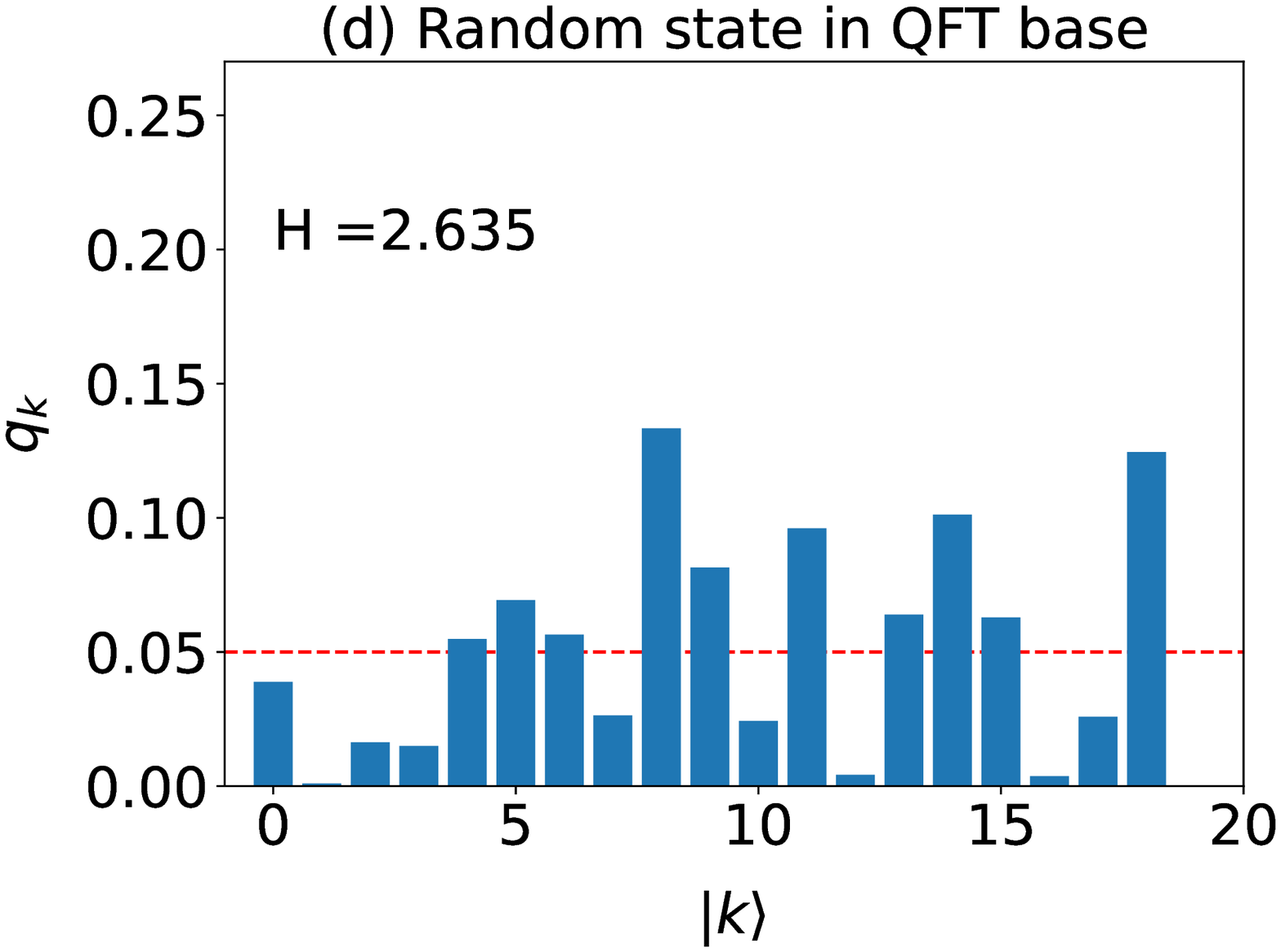}\\[10pt]
\includegraphics[width=0.48\textwidth]{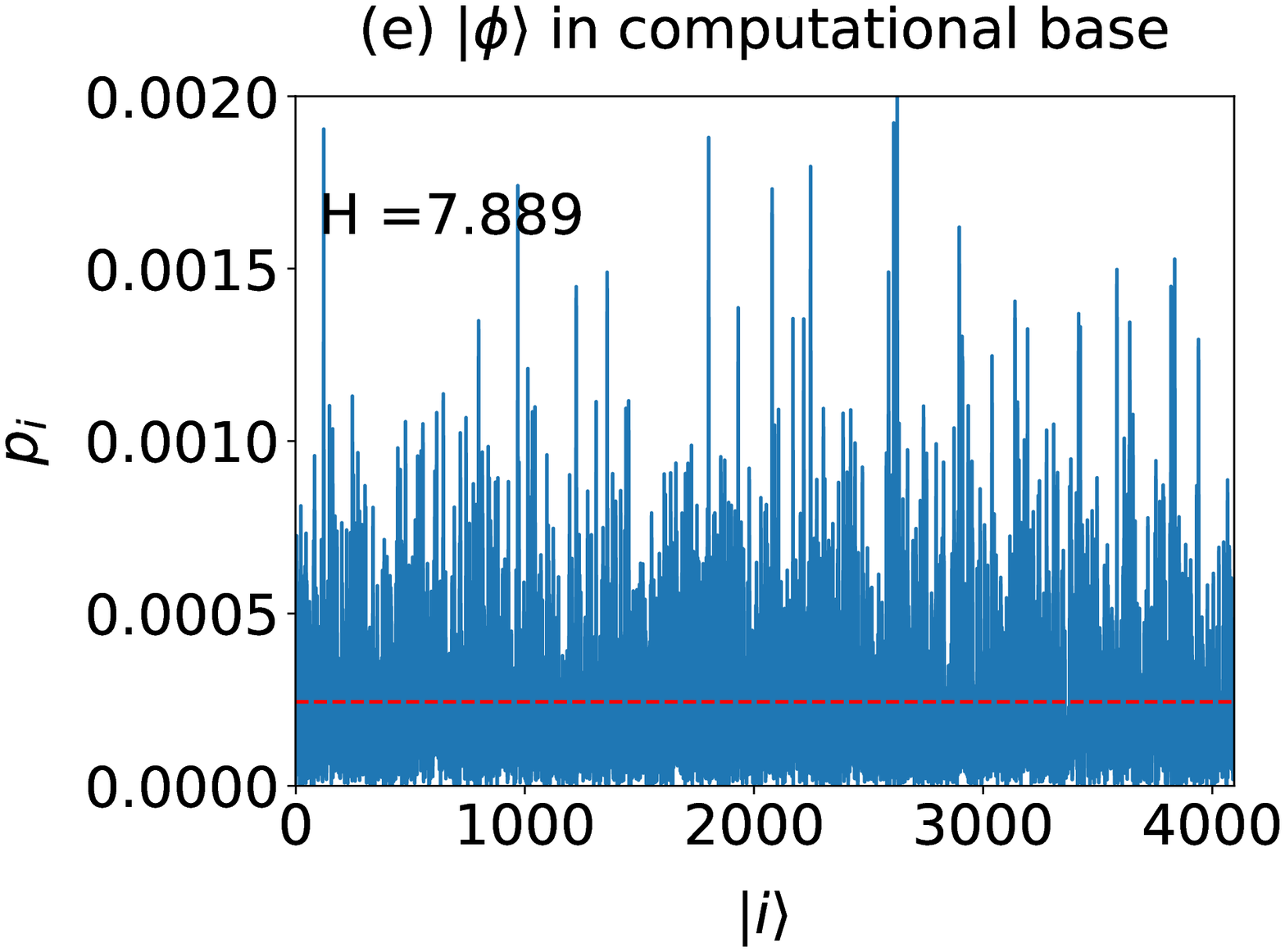}\quad
\includegraphics[width=0.48\textwidth]{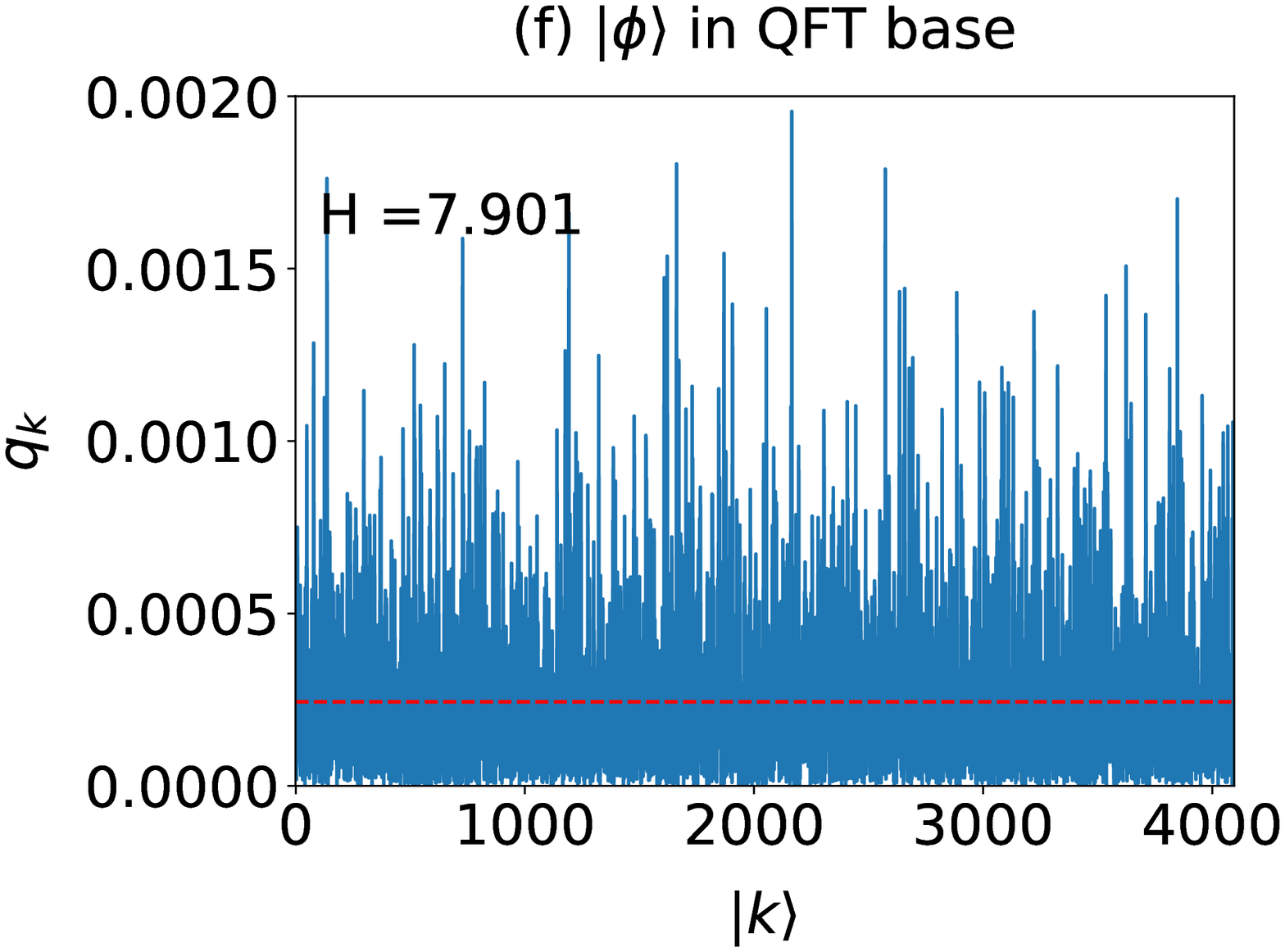}
}
\caption{The distributions of $p_i$ for (a) a uniform superposition of all basis states $N=20$, (c) a random 
quantum state generated by a random unitary matrix ${\rm U}(N)$ with $N=20$, and (e) a random quantum state 
of $n=14$ qubits generated by a random quantum circuit implemented on the Sycamore processor~\cite{Arute2019,
Martinis2022}. Plots (b), (d), and (f) are the distributions of $p_k$ after the quantum Fourier transform of 
(a), (b), and (c), respectively. The red lines indicate $p=1/N$ and the Shannon entropy $H$ for each state is 
labeled by $H$.
}
\label{Fig1}
\end{figure}

Note that the Shannon entropy $H(\ket{\psi})$ is defined with respect to a specific basis set $\{\ket{i}\}$. 
If the same quantum state is expanded in terms of  another basis set $\{\ket{k}\}$,
$\ket{\psi} = \sum_{k=0}^{N-1} b_k\ket{k}$, its Shannon entropy $H(\ket{\psi})$ 
will change. For two orthonormal basis sets, $\{\ket{i}\}$ and $\{\ket{k}\}$,
the entropic uncertainty relation~\cite{Deutsch1983,Kraus1987,MU1988,Coles2017} is written as
\begin{equation}
H(P) + H(Q) \ge -2\ln c 
\end{equation}
where $c=\max|\langle i|k\rangle|$, $P = (p_0,\dots,p_{N-1})$ with $p_i = |a_i|^2$, and
$Q = (q_0,\dots,q_{N-1})$ with $q_k = |b_k|^2$.
We consider the computational basis and the quantum Fourier transformed basis, which are mutually unbiased.
The quantum Fourier transform (QFT) is the discrete Fourier transform of the amplitude vector 
$(x_0,x_1,\dots, x_{N-1})$ to another amplitude vector $(y_0,y_1,\dots,y_{N-1})$, defined by
\begin{equation}
y_k = \frac{1}{\sqrt{N}}\sum_{j=0}^{N-1} e^{i2\pi\frac{jk}{N}}\, x_j\,.
\end{equation}
The QFT acting on $\ket{\psi_P}$ is written as
\begin{equation}
\ket{\psi_P} = \sum_{i=0}^{N-1}x_i\ket{i} \quad \longrightarrow\quad 
\ket{\psi_Q} =\sum_{k=0}^{N-1} y_k \ket{k} \,.
\end{equation}

Fig.~\ref{Fig1} illustrates the distributions of probabilities $p_i$ for two kinds of quantum states 
in the computational basis and probabilities $q_k$ in the QFT basis. First, consider a uniform superposition 
of all possible basis states in the computational basis that is given by 
$\ket{\psi} = \frac{1}{\sqrt{N}}(\ket{0} + \ket{1} + \cdots + \ket{N-1})$.  
As shown in Fig.~\ref{Fig1}~{(a)}, its $p_i$ is distributed uniformly, $p_i = {1}/{N}$ with $i=0,1,\dots,N-1$ 
and its Shannon entropy has the maximum value, $H(P) = \ln(N)$. 
The equally-likely distribution among $N$ possible states implies the largest randomness, and the greatest 
entropy~\cite{Ambegaokar1999}. However, $\ket{\psi}$ in the QFT basis set is localized at one site, so its 
entropy $H(Q)$ is zero, as depicted in Fig.~\ref{Fig1}~{(d)}. The entropic uncertainty is given by
$H(P) + H(Q) = \ln (N)$. Next, consider a random quantum state 
$\ket{\phi} = U_{\rm Haar}\ket{0}$. Fig.~\ref{Fig1}~{(b)} plots the distribution of $p_i$ of a random 
quantum state where $U_{\rm Haar}$ is generated by the QR decomposition of an $N\times N$ complex Gaussian 
random matrix. The entropy of a random quantum state is approximately given by 
$H(P)= \ln N -1+ \gamma\approx 2.302$ for $N=10$. Fig.~\ref{Fig1}~{(e)} plots the distribution of $q_k$ 
in the QFT basis.  Interestingly, {\it the entropy $H(P)$ of a random quantum state in the computational 
basis is almost equal to its entropy $H(Q)$ in its QFT basis.} We observe that the average entropy for 
all random quantum states in the computational basis, generated by a random circuit $U$, is equal to 
that in the QFT basis. A random quantum state may have the balanced entropic uncertainty, 
$H(P) = H(Q) = \ln N -1 + \gamma$. It is analogous to a coherent state in the sense that the latter has 
balanced or symmetric minimum uncertainty: $\Delta x = \Delta p$. It may be interesting to understand
why a random quantum state is invariant under the QFT. Fig.~\ref{Fig1}~{(c)} depicts the distribution of 
$p_i$ for a random quantum state generated by a random quantum circuit implemented on the Sycamore processor 
for $n=12$ and the cycles $m=14$~\cite{Martinis2022}. 
Fig.~\ref{Fig1}~{(f)} plots the distribution of $q_k$ in the QFT basis.
One can see that the Shannon entropy of a random quantum state in the computational basis  
is almost same as that of its QFT state. Note that the random circuit here is implemented on a classical 
computer without any noise. The Shannon entropy in Fig.~\ref{Fig1}~{e} is close to the theoretical 
value, $H=\ln N -1+\gamma \approx 7.8949$. The Shannon entropy calculated from the Sycamore data for $n=12$
and $m=14$ is $H\approx 8.217$ and close to $\ln(4096)\approx 8.31$~\cite{Oh2023b}.

\begin{figure}[t]
\centering{
\includegraphics[width=0.48\textwidth]{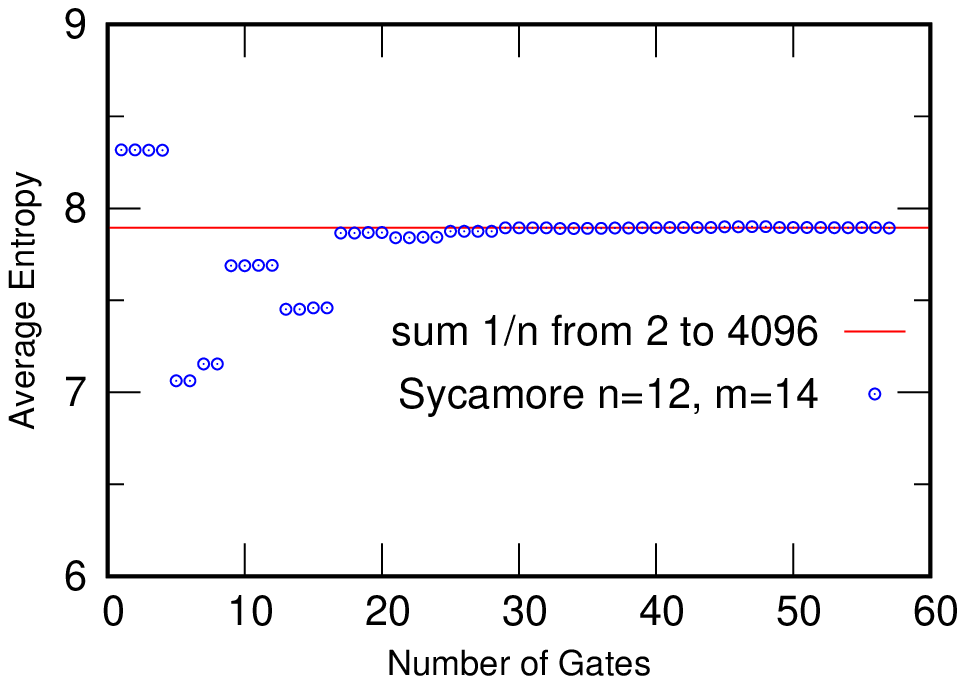}\quad
\includegraphics[width=0.48\textwidth]{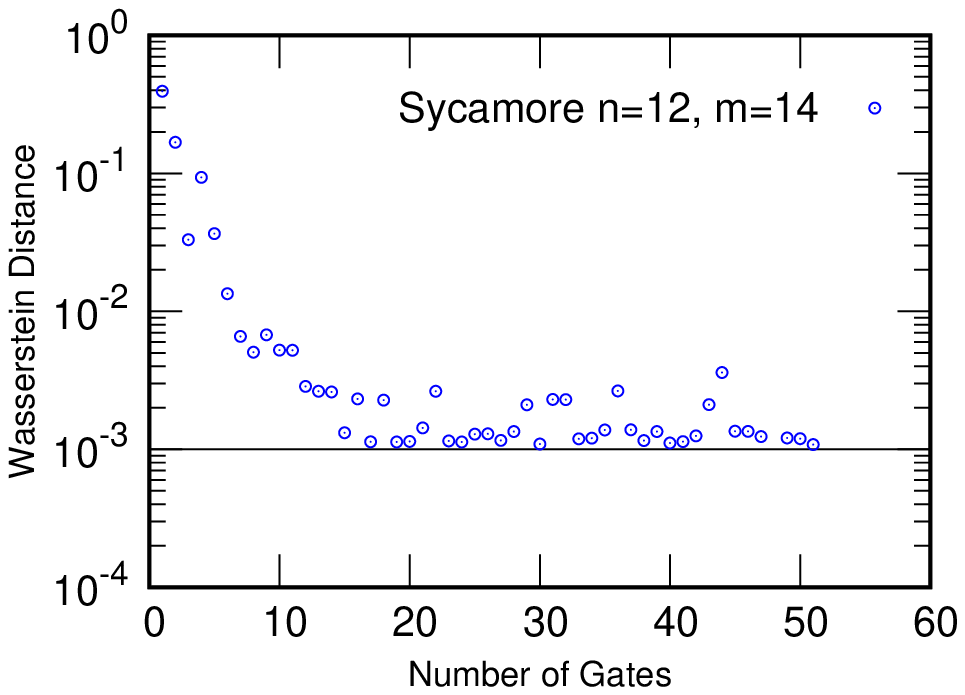}
}
\caption{For a random quantum circuit for $n=14$ qubits and up to $m=14$ cycles~\cite{Arute2019}, 
(a) Shannon entropy of the average of quantum states and (b) Wasserstein distance between the eigenvalues 
of random circuits and the Haar random unitary operator are plotted as a function of the number of gates applied,
i.e., the depth of a quantum circuit.}
\label{Fig2}
\end{figure}

Since a random quantum state is characterized by its Shannon entropy, $H(P) \approx \ln N -(1-\gamma)$,
the Shannon entropy $H(\ket{\psi^{(k)}})$ of a quantum state $\ket{\psi^{(k)}}=U_k\dots U_1\ket{0}$ 
at the $k$-th step of random walks could be used as a convergence measure to see whether the cutoff 
phenomenon happens.
The sub-linear convergence proportional to $\sqrt{n}$ was claimed by calculating the average entropy of 
quantum states~\cite{Boixo2018}. As shown in Fig.~\ref{Fig2}~{(a)}, the Shannon entropy of a quantum
state converges to $\ln N -(1-\gamma)$ as the number of gates increases and remains there.
An interesting question is what happens to a quantum state after the Shannon entropy converges.

The eigenvalue distribution for random unitary operators drawn from the Haar measure is well 
known~\cite{Meckes2019}, so the distance between the eigenvalue distributions could be used to measure
the convergence of a random walk. We consider the Wasserstein distance defined by 
\begin{equation}
W_p(P,Q) = \left( \inf_{J\in J(P,Q)} \int ||x-y||^p\, dJ(x,y) \right)^{1/p}\,,
\end{equation}
where ${\cal J}(P,Q)$ denotes all joint distributions $J(P,Q)$. If $P$ and $Q$ are the empirical distributions 
of a data set $\{x_1,\dots,x_n\}$ and $\{y_1,\dots,y_n\}$ 
respectively, then the Wasserstein distance is given by the distance between order statistics
\begin{equation}
W_p(P,Q) =\left( \sum_{i=1}^n || x_{(i)} -y_{(i)}||^p \right)^{1/p}\,,
\end{equation}
where $x_{(i)}$ is the $i$-th order statistic of a samples, i.e., its $i$-th smallest value. 
Fig.~\ref{Fig2}~{(b)} plots the Wasserstein distance of order 1 between the eigenvalue distribution
of the circular unitary ensemble and that of a random quantum circuit as a function of the number of
quantum gates applied. The calculation of the Wasserstein distance supports the cutoff phenomenon
for random quantum circuits as the Shannon entropy does.

\section{Dyson-Brownian Motions on a Unitary Group \label{Sec3}}

In Sec.~\ref{Sec2}, the random walk on the unitary group is implemented by applying the sequence of
random quantum circuits, $\{U_1,U_2,\dots\}$. A random quantum state after the $k$-th step is 
$\ket{\psi} = U_k\dots U_1\ket{0}$. This may be considered as a discrete process.
In this section, we consider a continuous random walk given by a time-dependent random Hamiltonian  
to see how quickly a quantum state converges to a random quantum state.
The time evolution operator at $t + \delta t$ is given by
\begin{align}
U(t+\delta t) &= e^{i\frac{\delta t}{\hbar} H(t)}\, U(t)\\
        &\approx  \Bigl[1 + i\frac{\delta t}{\hbar} H(t) -\frac{1}{2}\left(\frac{\delta t}{\hbar}\right)^2 H^2(t) 
        \Bigr] U(t) \,,
\label{Evolution}
\end{align}
where $U(0) = I$. The time dependent random Hamiltonian $H(t)$ at time $t$ is obtained as follows. 
We draw a complex random matrix $A$ whose real and imaginary parts of a matrix element $A_{ij}$ 
are sampled independently from the normal distribution ${\cal N}(0,\sigma^2/2N)$ with the variance $\sigma^2/2N$. 
The Hermitian property of the Hamiltonian $H$ is fulfilled by summing $A$ and its conjugate transpose 
$A^\dag$, $H = \frac{1}{2}(A + A^\dag)$. $H$ is the elements of the Gaussian orthogonal ensemble.

\begin{figure}[t]
\centering{
\includegraphics[width=0.48\textwidth]{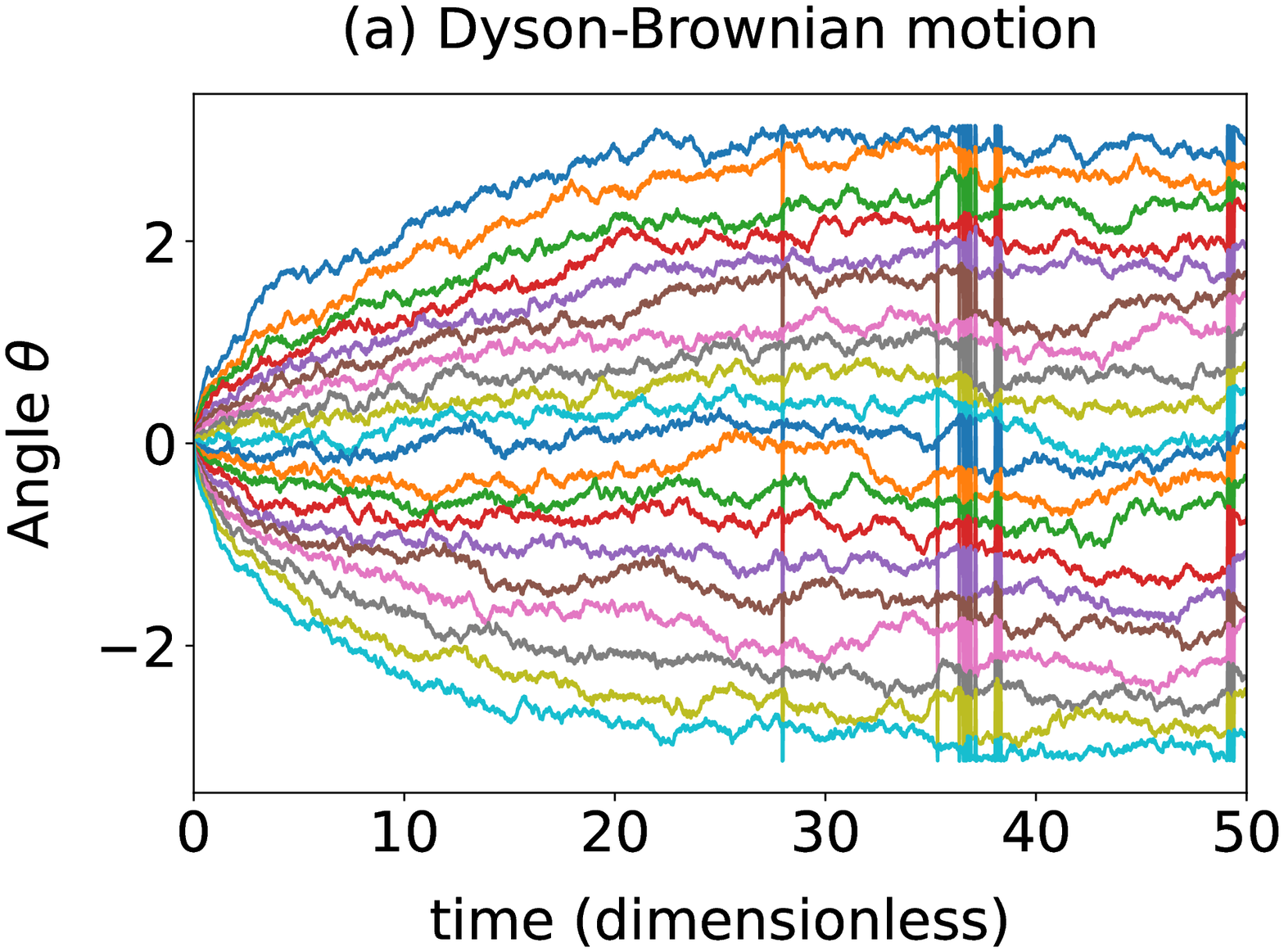}\quad
\includegraphics[width=0.48\textwidth]{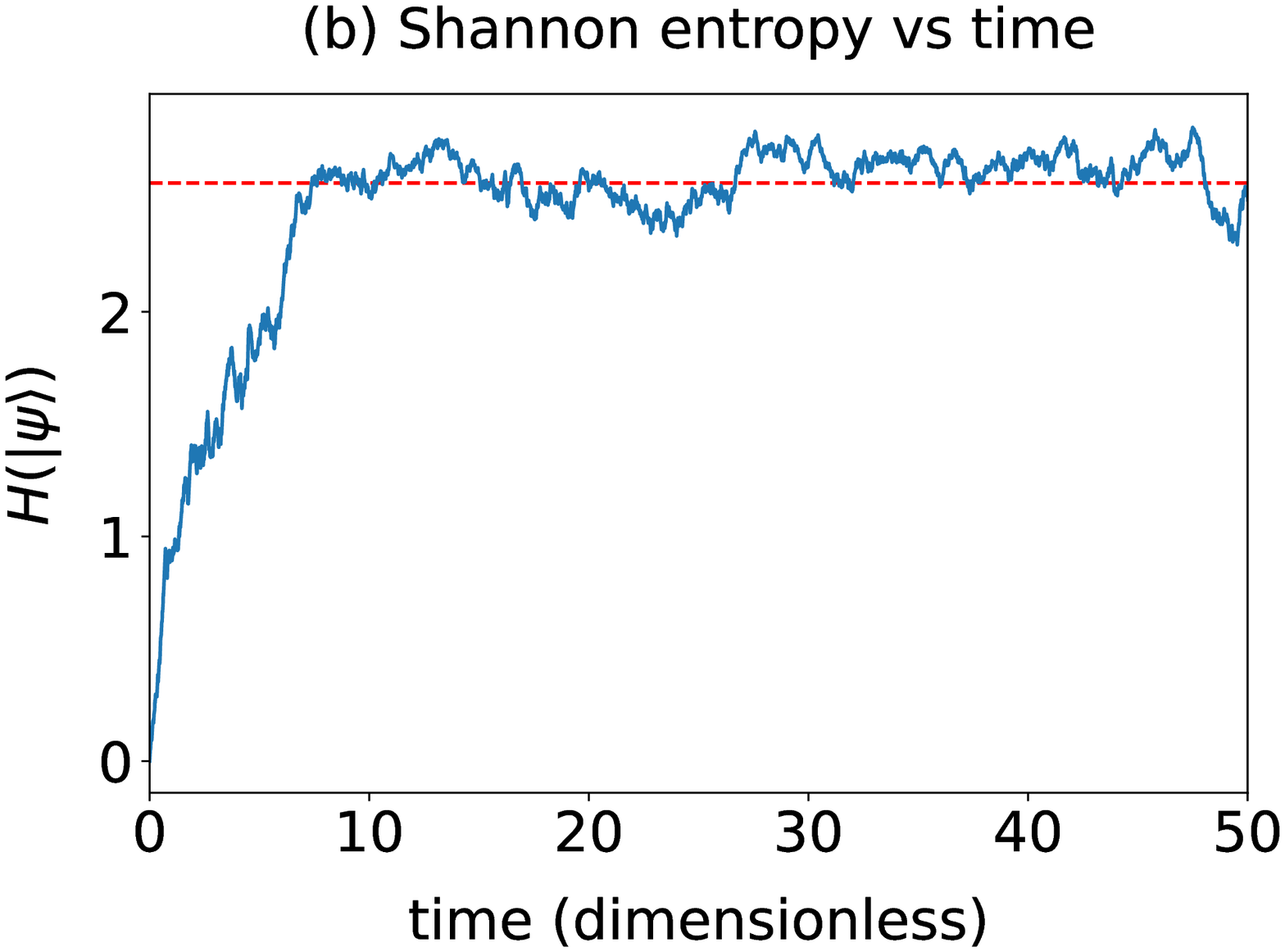}
}
\caption{For the Dyson-Brownian random walk, (a) trajectories of eigenvalues of a random unitary operator and 
(b) the Shannon entropy $H(\ket{\psi})$ of a quantum state are plotted. 
Here, we take $20\times 20$ random Hamiltonian matrices,
i.e., $N = 20$ and the time step $\delta t = 0.01$. In (b) the red dotted horizontal line represents
the Shannon entropy of a random quantum state, $\ln N -1 + \gamma \approx 2.5572$ for $N=20$.
}
\label{Fig3}
\end{figure}

The trajectories of eigenvalues of a random unitary operator $U$ are known as the Dyson-Brownian motions 
and do not overlap with others. Eigenvalues repel each other. We simulate the time evolution with 
a time-dependent random Hamiltonian for $N=20$ and $\sigma=1$ so $\sigma^2/2N=1/2N$. For simplicity, 
we take $\hbar = 1$. We set the time step $\delta t=0.01$. The eigenvalues of $U(t)$ from 
Eq.~(\ref{Evolution}) are obtained by diagonalizing it and their trajectories as a function of time 
are plotted in Fig.~\ref{Fig3}~{(a)}. The time-evolution of the Shannon entropy of a quantum state 
$\ket{\psi(t)} = U(t)\ket{0}$ is shown in Fig.~\ref{Fig3} (b). We observe the non-asymptotic convergence 
to $\ln N -1 +\gamma\approx 2.5572$ for $N=20$.  
It is interesting to see the eigenvalues and the Shannon entropy fluctuate after 
arriving at the steady state. Particularly, the fluctuation in the Shannon entropy in Fig.~\ref{Fig3} (b) 
is in contrast to no fluctuation in the Shannon entropy for a finite random walk of a random quantum 
circuit, shown in Fig.~\ref{Fig1}~{(a)}.


\section{Summary}
\label{Sec_summary}
In this paper, we examined some properties of random quantum states generated by discrete and continuous 
random walks on a unitary group. It is found that the Shannon entropy of a random quantum state generated 
by random quantum circuits is invariant under the QFT in the sense that the Shannon entropy does not change 
before and after applying the QFT. The entropic uncertainty relation of a random quantum state for 
the computational bases and the QFT bases is balanced, i.e., $H(P)=H(Q)$. This may remind us of the coherent state 
with the balanced minimum uncertainty relation, $\delta x = \delta p$. We showed that the cutoff 
phenomenon for a random quantum circuit occurs by calculating the Shannon entropy and the Wasserstein distance 
for the eigenvalue distributions. It is a open question whether the cutoff of random walks on
a unitary group scales with the number of qubits $n$ as sub-linear ${\cal O}(\sqrt{n})$ or ${\cal O}({n} 2^n)$ 
while the numerical calculations here seem to support the sub-linear scaling of the cutoff.

In addition to the demonstration of quantum advantage, random quantum circuits may be applicable to solving
interesting problems, for example, in randomized linear algebra. The set of random quantum states 
$\ket{\psi_i} = U_{\rm Haar}\ket{i}$ with $i=0,1,\dots,2^n-1$ form the orthonormal basis set. 
The trace of a matrix $A$ could be calculated using 
${\rm Tr}(A) \approx \frac{1}{M}\sum_{i=1}^M \bra{\psi_i}A\ket{\psi_i}$ where $\ket{\psi_i}$ is 
a random quantum state~\cite{Oh2023}. 

\section*{Acknowledgment}
This material is based upon work supported by the U.S. Department of Energy, Office of Science, 
National Quantum Information Science Research Centers.
We also acknowledge the National Science Foundation under award number 1955907. 

\section*{References}
\bibliographystyle{iopart-num}
\bibliography{Paper_Oh_Kais}
\end{document}